# Extracting the sample response function from experimental two-dimensional terahertz-infrared-visible spectra

Pankaj Seliya[1], Mischa Bonn[1], and Maksim Grechko[1]*

[1]Department of Molecular Spectroscopy, Max Planck Institute for Polymer Research, Ackermannweg 10, D-55128, Mainz, Germany

*Corresponding author

Email: grechko@mpip-mainz.mpg.de

## Abstract

Terahertz molecular motions are often probed by high-frequency molecular oscillators in different types of non-linear vibrational spectroscopy. Recently developed two-dimensional terahertz-infrared-visible spectroscopy allows direct measuring of this coupling and, thus, obtaining site-specific terahertz vibrational spectrum. However, these data are affected by the intensity and phase of the employed laser pulses. In this work, we develop a method of extracting sample response – representing solely physical properties of a material - from experimental spectra. Using dimethyl sulfoxide (DMSO) as a model molecule to verify this method, we measure the coupling between C-H stretch vibration of its methyl groups and terahertz intramolecular twist and wagging modes.

## Introduction

Molecular low-frequency vibrational modes (LFMs) in the terahertz (THz) frequency range are essential for (bio-) chemical processes in condensed phases at room temperature [1–4]. Non-linear mid-infrared (MIR) spectroscopy techniques (transient absorption and 2D IR) measure high-frequency vibrational modes (HFMs) to obtain insight into these motions at locations of specific chemical groups [5–10]. Such measurements probe terahertz dynamics indirectly via LFM-HFM coupling, by exciting and probing only a high-frequency oscillator. Sophisticated relations between the HFM frequency/transition dipole moment and the LFM coordinate complicate determining the relevant LFM modes from such measurements. Recently, we have developed two-dimensional terahertz-infrared-visible (2D TIRV) spectroscopy to measure LFM-HFM coupling directly [11,12].

This technique can identify the relevant LFMs using mid-infrared spectroscopy selectivity, thereby disentangling the typically broad featureless low-frequency spectral response [13–16]. However, taking full advantage of the 2D TIRV spectroscopy requires extracting the system's complex-valued third-order non-linear response function $S^{(3)}$ from the measured data.

A measured 2D TIRV spectrum is determined by a product of $S^{(3)}$ and spectra of laser pulses. The latter can be considered as an instrument response function (IRF) which affects both the intensity and phase of the 2D spectrum. Deriving $S^{(3)}$, which contains the physical properties of a sample, thus requires eliminating IRF from the experimental data. This requirement is analogous to data processing in other non-linear spectroscopy techniques where the signal depends on the intensities and phases of the employed laser pulses. For example, progress in 2D IR and 2D electronic spectroscopy was enabled by a method of determining the phase of a spectrum [17–21]. The method utilizes the equality of 2D spectrum projection and transient absorption spectrum. Another technique, phase-resolved sum-frequency generation (PR SFG) spectroscopy, uses the response of reference material to obtain the second-order response function of a sample [22,23]. Here, we develop a similar method for 2D TIRV spectroscopy.

We have recently reported a procedure that allows separating different excitation pathways in 2D TIRV spectroscopy [24]. This was a critical step towards deriving material response function from experimental data, because different excitation pathways produce signals via different mixing of the terahertz and infrared fields - sum- and difference-frequency mixing. Here, we report a method to eliminate the effect of the laser fields and obtain the spectrum of sample response function in 2D TIRV spectroscopy. To this end, we utilize a thin (1 μm) silicon nitride (SiN$_x$) membrane as reference material. Its response gives a spectrum that directly reflects the distribution of intensities and phases of all the laser pulses used in 2D TIRV spectroscopy to generate and detect signal fields. We demonstrate the approach on a model sample - liquid dimethyl sulfoxide (DMSO), which has intense low- and high-frequency modes that are sufficiently strongly coupled to generate a substantial 2D TIRV signal.

## Principles of 2D TIRV spectroscopy

The theoretical formalism and experimental implementation of 2D TIRV spectroscopy have been discussed in detail elsewhere [11,24,25]. In brief, in a 2D TIRV spectroscopy measurement, four-wave mixing of a narrowband visible (VIS), and broadband terahertz (THz) and mid-infrared (IR) pulses induces the emission of a signal field by a sample. The signal field is measured using its interference with a local oscillator (LO) pulse at different time delays between the THz pulse and IR / VIS pulse pair. Fourier transformation of the time-domain data provides a spectrum. Assuming an infinitely small bandwidth of the VIS pulse, a measured 2D TIRV spectrum $\Gamma$ is given by a product:

$$\Gamma(\Omega_2 + \Omega_{Vis}, \Omega_1) \propto \{S^{(3)}(\Omega_2 + \Omega_{Vis}, \Omega_2, \Omega_1) + S^{(3)}(\Omega_2 + \Omega_{Vis}, \Omega_2, \Omega_2 - \Omega_1)\} \\ \times E_{THz}(\Omega_1) E_{IR}(\Omega_2 - \Omega_1) E_{VIS}(\Omega_{VIS}) E_{LO}(-\Omega_2 - \Omega_{Vis}) \quad (1)$$

Here $S^{(3)}$ is the third-order response function of a sample, which reflects its physical properties; $E_{VIS}$ is the complex-valued spectral amplitude of the VIS electric field with frequency $\Omega_{VIS}$; $E_{THz}$, $E_{IR}$ and $E_{LO}$ are complex-valued electric field spectra of THz, IR, and LO pulses, respectively. The first and second terms in Eq. 1 represent different interaction sequences between the THz and IR fields and the sample. In the first (second) term, the interaction of the sample with THz (IR) precedes interaction with IR (THz). Equation 1 shows that the intensity and phase of a measured 2D TIRV spectrum are affected by the spectral shapes and phases of the employed laser pulses. We note that the THz, IR, and LO pulse phases can have complex dependence on frequency (due to, e.g., dispersion). Hence, a measured spectrum can be considered a product of the sample response function and instrument response function (IRF), which is given by the product of electric fields.

Deriving the complex-valued spectrum of $S^{(3)}_{sample}$ requires eliminating laser fields from the product in Eq. 1. To this end, we use an approach similar to PR SFG spectroscopy [22]. In this technique, a measured spectrum $\Gamma_{sample}$ of a sample is divided by a non-resonant spectrum $\Gamma_{ref}$ of a reference material (typically quartz). For non-resonant interactions, a material response function of any order is real-valued and constant (frequency-independent). Thus, the non-resonant spectrum is linearly proportional to the IRF of an experimental setup, and $\Gamma_{sample}/\Gamma_{ref} \propto S^{(3)}_{sample}$. This approach can be employed using a resonant reference material as well, given that its response function $S^{(3)}_{ref}$ is known. In this case:

$$S^{(3)}_{sample} = \frac{\Gamma_{sample}}{\Gamma_{ref}} S^{(3)}_{ref}$$

We demonstrate that in 2D TIRV spectroscopy, a thin silicon nitride ($SiN_x$) membrane is a suitable reference material to derive the complex-valued spectrum of the sample response function.

## Results and Discussion

We use liquid dimethyl sulfoxide (DMSO) as a model sample. The absorption spectrum of DSMO in the mid- through far-infrared frequency range is shown in Fig. 1. In the terahertz frequency range, this spectrum contains two sharp vibrational modes at 333 cm$^{-1}$ and 383 cm$^{-1}$ assigned to intramolecular CH$_3$ - CH$_3$ twist ($v_{23}$) and wagging ($v_{11}$) modes, respectively [26–28]. There is also a weak shoulder at about 308 cm$^{-1}$ ($v_{12}$), which is also produced by intramolecular motion and is assigned to C-S-C bending [28]. Besides intramolecular vibrations, the spectrum contains intermolecular libration motion at frequencies below 150 cm$^{-1}$ [26].

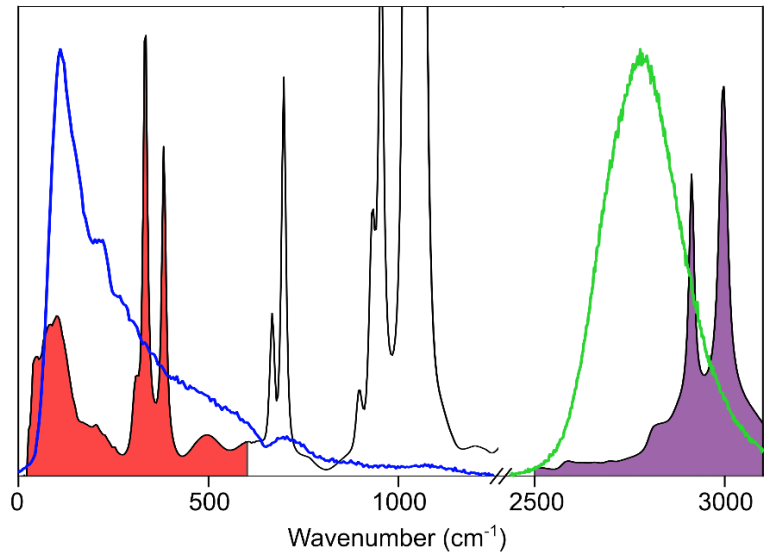

Figure 1. The black line shows the absorption spectrum of liquid DMSO. Relevant low- and high-frequency ranges are shaded in red and purple, respectively. Intensity spectra of THz and IR laser pulses are shown by blue and green lines, respectively.

Because the 333 cm$^{-1}$ and 383 cm$^{-1}$ LFMs of DMSO displace CH$_3$ groups of the molecule, one can expect substantial coupling between LFMs and the C-H stretch vibrations. The C-H stretch symmetric and asymmetric vibrational modes have frequencies of 2913 cm$^{-1}$ and 2997 cm$^{-1}$, respectively (Fig. 1). Figure 2a shows the first and second quadrants of the absolute-value 2D TIRV spectrum of DMSO ($|\Gamma_{\text{DMSO}}|$) in this IR frequency range, which is measured using a femtosecond IR pulse with a bandwidth of 250 cm$^{-1}$, and a central frequency of 2780 cm$^{-1}$ (Fig. 1). The corresponding time-domain data are shown in Fig. 2b. Sum-frequency mixing of the THz and IR fields produces a signal in the first quadrant at $\omega_2$-frequencies of CH$_3$ stretch vibrations (2913 cm$^{-1}$ and 2996 cm$^{-1}$) [24]. The offset of the IR pulse frequency from CH$_3$ modes to red favors this process. The similar signal in the second quadrant is weak because it is produced by difference-frequency mixing of the IR and THz fields, which is hampered by the IR frequency offset. Instead, the second quadrant contains signal parallel to the diagonal and intersecting $\omega_2$-frequency axis at the CH$_3$ stretch frequencies.

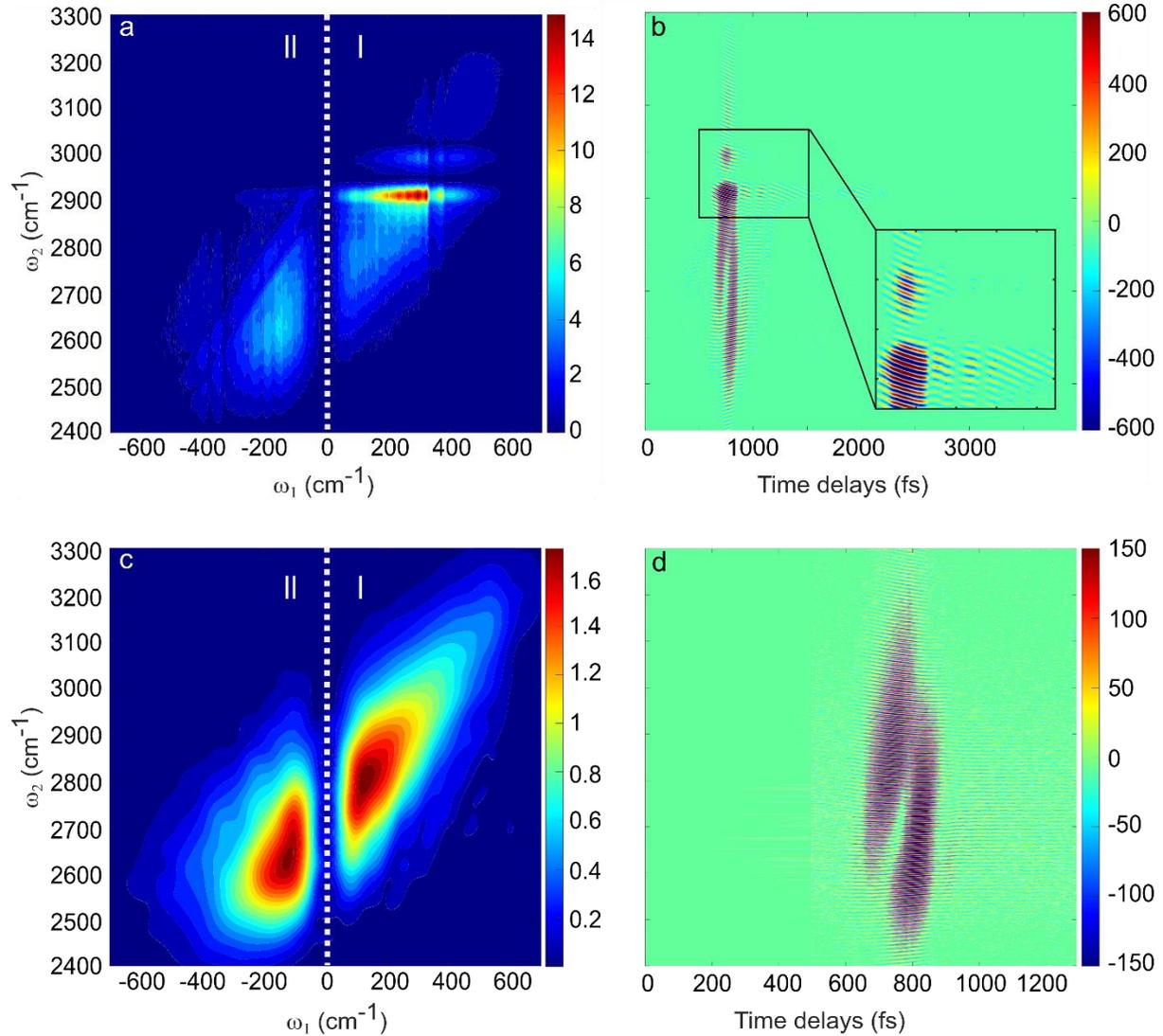

Figure 2. Absolute value 2D TIRV spectrum and time domain data of DMSO (a,b) and 1 μm thick SiN$_x$ membrane, which serves as reference (c,d). In panels (b,d), the time zero is arbitrarily defined as the beginning of a data set.

The shape of the $\Gamma_{\text{DMSO}}$ is affected by the spectral intensities and phases of the THz, IR, and LO pulses. To eliminate this effect, we measure the 2D TIRV spectrum $\Gamma_{\text{SiN}}$ of a reference material, a ~1 μm thick SiN$_x$ membrane (Figs. 2c,d). Because SiN$_x$ has no resonances at mid-infrared and visible frequencies, interaction with IR and VIS pulses is non-resonant for this material. In contrast, in the far-infrared frequency range, SiN$_x$ has intense vibrational resonances (Fig. 3). Hence, the signal can be generated by its non-resonant as well as resonant interaction with the THz pulse. In the former case SiN$_x$ response function $S_{SiN}^{(3)}$ is real-valued and constant, whereas in the latter case it is complex-valued and depends on the $\omega_1$-frequency:

$$S_{SiN}^{(3)}(\omega_2, \omega_1) \propto S_{SiN}^{(1)}(\omega_1)$$

Here $S_{SiN}^{(1)}$ is the resonant first-order response function of SiN$_x$. Figure 3 shows $S_{SiN}^{(1)}$ spectrum calculated using its reported permittivity $\varepsilon_{SiN}$ [29]:

$$S_{SiN}^{(1)}(\omega) = \frac{\varepsilon_{SiN}(\omega) - 1}{4\pi\rho_0},$$

where $\rho_0$ is the number density of absorbers (we assume $\rho_0 = 1$). In the 0-700 cm$^{-1}$ terahertz frequency range of our 2D TIRV experiment, the $S_{SiN}^{(1)}$ amplitude varies by about 67%, and its phase by up to 0.8 rad. This dispersion is primarily caused by the transverse optical phonon resonance around 825 cm$^{-1}$.

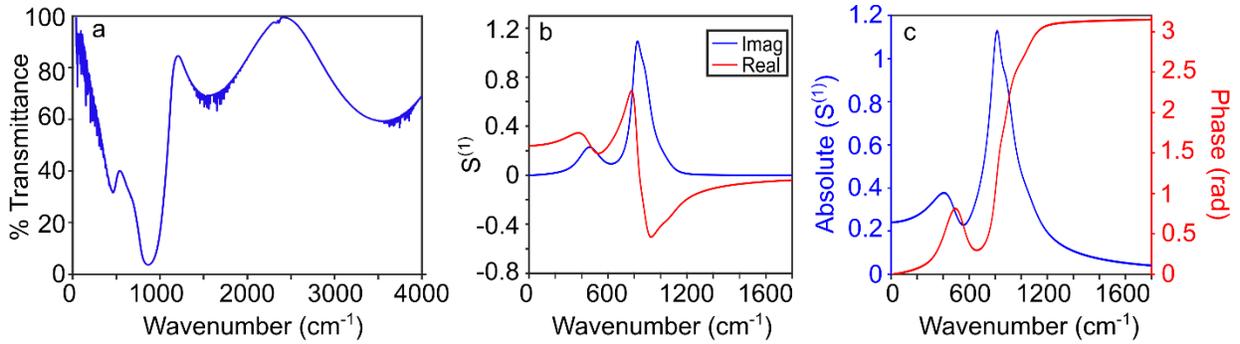

Figure 3. (a) Transmittance spectrum of 1 µm thick SiN$_x$ membrane. (b) Real and imaginary parts, (c) amplitude and phase of $S_{SiN}^{(1)}$.

Using SiN$_x$ as a reference material, we test the two possible extreme cases of interaction with the THz pulse. Figures 4(a-c) and (d-f) show $S_{DMSO}^{(3)}$ in the first quadrant obtained assuming fully non-resonant (case I) and partly resonant (for the THz pulse interaction, case II) SiN$_x$ signal generation, respectively:

$$(\text{a-c}) \quad S_{DMSO}^{(3)}(\omega_2, \omega_1) = \frac{\Gamma_{DMSO}(\omega_2, \omega_1)}{\Gamma_{SiNx}(\omega_2, \omega_1)},$$

$$(\text{d-f}) \quad S_{DMSO}^{(3)}(\omega_2, \omega_1) = \frac{\Gamma_{DMSO}(\omega_2, \omega_1)}{\Gamma_{SiNx}(\omega_2, \omega_1)} \cdot S_{SiN}^{(1)}(\omega_1).$$

**Case I: Fully non-resonant SiN$_x$ signal**

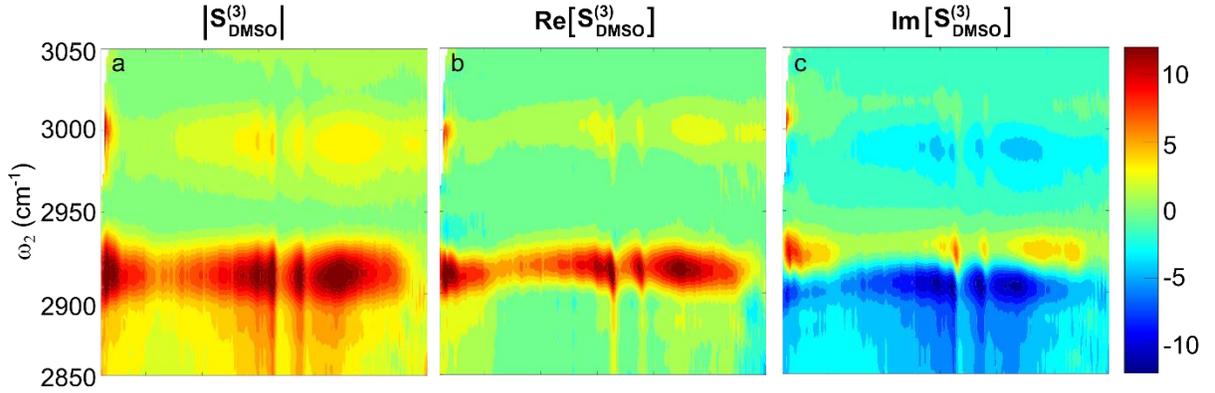

**Case II: Partly resonant SiN$_x$ signal**

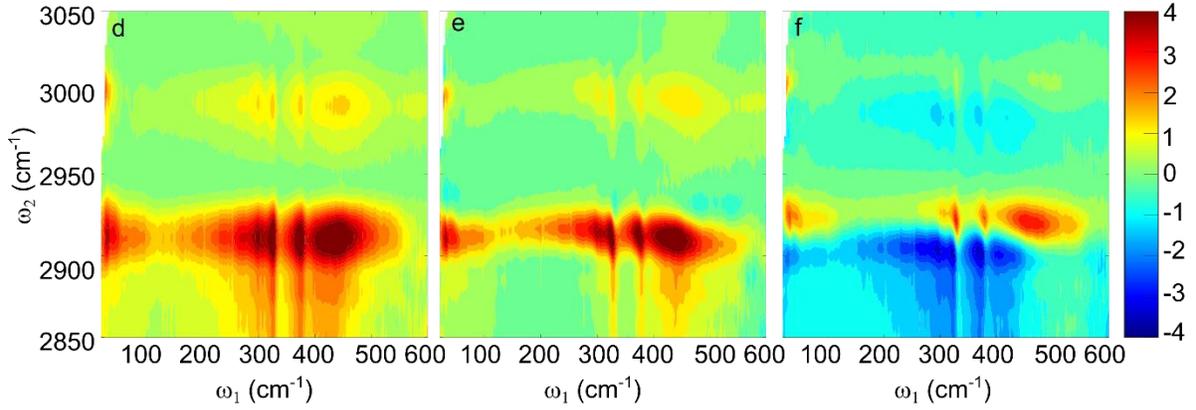

Figure 4. (a-c) Absolute value, along with the real and imaginary parts of DMSO response function spectrum assuming fully non–resonant SiN$_x$ signal; (d–f) Absolute value, along with the real and imaginary parts of DMSO response function spectrum, assuming partly resonant SiN$_x$ signal.

These two assumptions produce $S_{\text{DMSO}}^{(3)}$ spectra which are very similar for $\omega_1 \lesssim 400$ cm$^{-1}$. For higher $\omega_1$-frequencies, the difference of the lineshape becomes more significant because of the ~0.8 rad accumulated phase of $S_{\text{SiN}}^{(1)}$. A 2D spectrum with the correct phase should fulfill Kramers-Kronig relations calculated parallel to one of the frequency axis [30] for instance:

$$\text{Im}\left(S_{\text{DMSO}}^{(3)}(\omega_2, \omega_1)\right) = -\frac{1}{\pi} \int_{-\infty}^{+\infty} \frac{\text{Re}\left(S_{\text{DMSO}}^{(3)}(\omega_2, \omega')\right)}{\omega' - \omega_1} d\omega'$$

Figure 5 shows real and imaginary parts of the spectra in Fig. 4 taken at the symmetric stretch frequency ($\omega_2 = 2913$ cm$^{-1}$). For case II, the imaginary part inferred from the real part using the Kramers-Kronig relation (KKR) agrees reasonably with the experimental data (Fig. 5d), given the limited spectral range (0-600 cm$^{-1}$) used in the calculation and possible non-resonant contribution in the DMSO signal. For case I, the agreement between calculated and experimental spectra is significantly worse (Fig. 5b). Thus, the 2D TIRV signal of SiN$_x$ is generated predominantly via resonant interaction with the THz pulse, and the correct spectrum of $S^{(3)}_{DMSO}$ is given by case II.

**Case I: Fully non-resonant SiN$_x$ signal**

**Case II: Partly resonant SiN$_x$ signal**

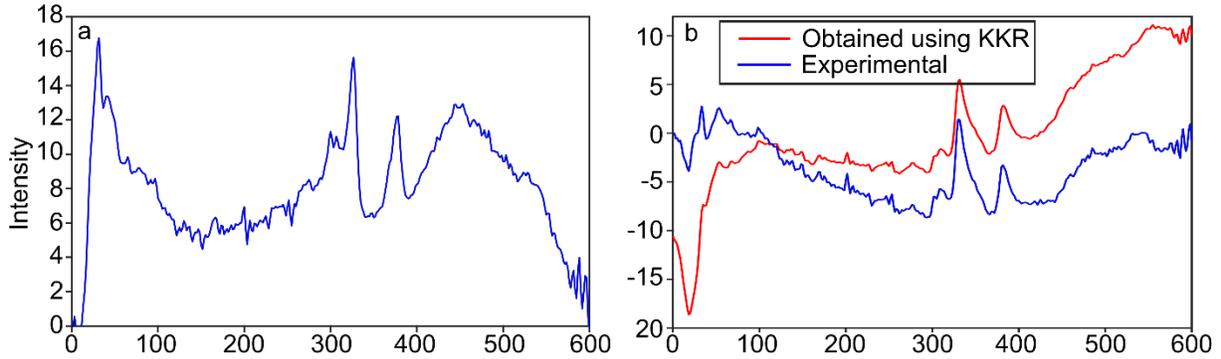
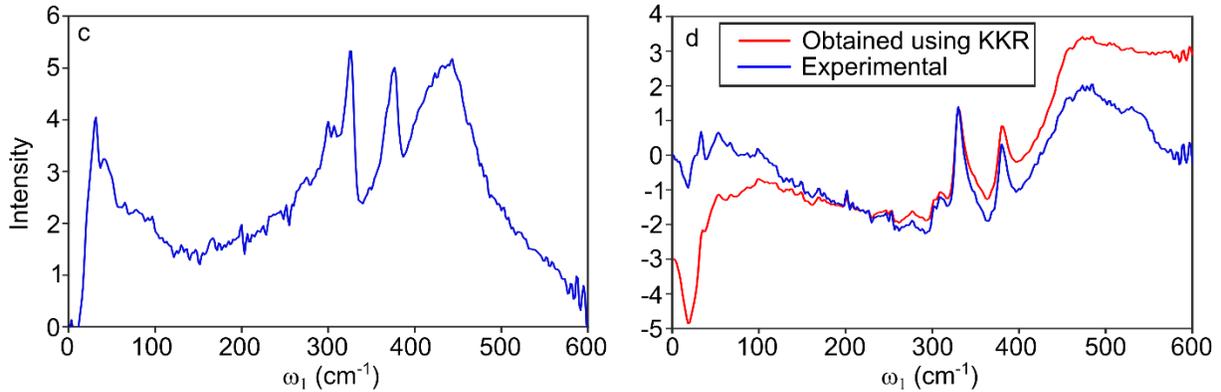

Figure 5. Real (a) and imaginary (b) parts of $S^{(3)}_{DMSO}(2913 \text{ cm}^{-1}, \omega_1)$, assuming fully non-resonant SiN$_x$ signal. Real (a) and imaginary (b) parts of $S^{(3)}_{DMSO}(2913 \text{ cm}^{-1}, \omega_1)$, assuming partly resonant SiN$_x$ signal. Blue lines show experimental data, red lines show the imaginary part calculated from the corresponding real part using the Kramers-Kronig relation.

For each of the C-H stretch vibrational modes, the $S^{(3)}_{DMSO}$ spectrum contains three peaks at $\omega_1 \approx$ 300 cm$^{-1}$, 326 cm$^{-1}$ and 376 cm$^{-1}$ reflecting their coupling with the intramolecular LFMs. To determine which of the THz-IR and IR-THz interaction sequences produces these signals, we perform a Fourier transformation of the spectrum over the $\omega_1$-frequency axis back into the time domain (Fig. 6a). Because the data processing eliminates phases of the laser pulses, $S^{(3)}_{DMSO}$ has maximum at $t = 0$ (note that here the $t$-axis is uniquely determined by the phase of the frequency-domain spectrum). We use a square window function to select data at negative ($t < 0$) and positive ($t > 0$) time and perform their Fourier transformation separately to derive the signal generated by the IR-THz and THz-IR interaction sequence, respectively. The spectrum of the signal at $t < 0$ contains peaks that are broad over the $\omega_1$ axis and narrow over the $\omega_2$ axis (Fig. 6(b-d)). Thus, the IR-THz interaction sequence is resonant for the IR interaction and presumably non-resonant for the THz interaction. In contrast, the spectrum of $t > 0$ contains a signal generated by the coupled intramolecular LFMs and C-H stretch vibrations (Figs. 6(e-g)). Hence, the doubly-resonant signal is produced by DMSO molecules interacting first with the THz pulse, followed by interaction with the IR pulse.

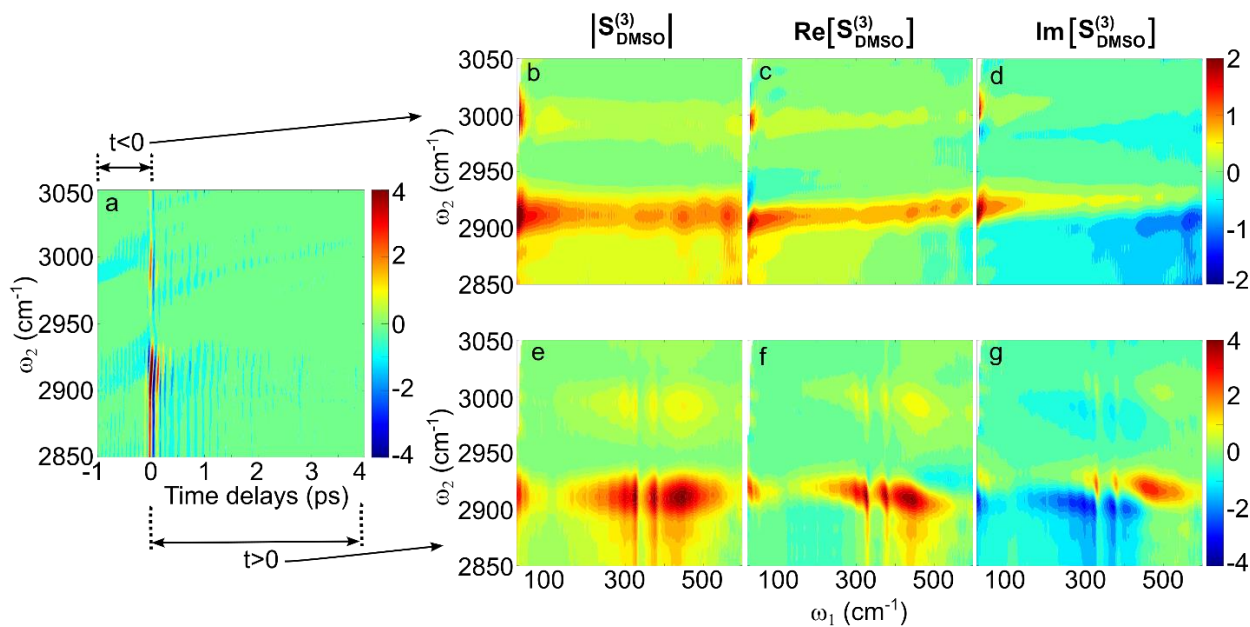

Figure 6. (a) Time domain data obtained by inverse Fourier transformation of $S^{(3)}_{\text{DMSO}}(\omega_2, \omega_1)$ over $\omega_1$ frequencies. (b) Absolute value, (c) real and (d) imaginary parts of the spectrum obtained by Fourier transformation of the signal for $t < 0$ in (a). (e) Absolute value, (f) real and (g) imaginary parts of the spectrum obtained by Fourier transformation of the signal for $t > 0$ in (a).

Cross-peaks between $CH_3$ stretch vibrations and intra-molecular LFMs are at frequencies of $\omega_1 \approx$ 300 cm$^{-1}$, 326 cm$^{-1}$ and 376 cm$^{-1}$. These frequencies are systematically lower by about 5 cm$^{-1}$ than LFMs locations in the absorption spectrum (accuracy of the $\omega_1$- and $\omega_2$-frequencies is about 1 cm$^{-1}$ each, see Supplementary Information for details). This deviation cannot be explained by different frequencies of LFMs in the excited state of the C-H oscillator because THz pulse excites LFMs in the ground vibrational state of C-H stretch. It also cannot be explained by own mechanical anharmonicity of the LFMs (see Supplementary Information for details). Hence, we tentatively attribute this red-shifted response to a slight inhomogeneous broadening of the LFMs. We speculate that coupling is stronger for the LFMs at the lower-frequency side of the spectrum [31].

At even lower $\omega_1$-frequency, the $S_{\text{DMSO}}^{(3)}$ spectra display coupling between C-H stretch vibration and intermolecular librational motions (Figs. 4,6). The librational motion changes the spatial orientation of the C-H stretch transition dipole moment, and because the intermolecular forces on the $CH_3$ groups are typically rather weak, this coupling is likely produced by predominantly electrical anharmonicity. The maximum intensity in the 2D spectrum is substantially below THz frequency of 100 cm$^{-1}$ – the maximum of linear absorption (Fig. 1). We hypothesize that this can be due to the larger amplitude (angle) of libration at lower frequencies and, thus, stronger LFM-HFM coupling.

In summary, we present a method of deriving complex-valued sample response function in experimental 2D TIRV spectroscopy. The sample response free from the instrument response function reflects its physical properties and can be directly compared with theoretical calculations. For a model liquid DMSO sample, we directly measure the coupling between high-frequency C-H stretch modes and THz molecular motions, both intramolecular vibrations and intermolecular libration. These results pave the way for the utilization of 2D TIRV spectroscopy for the quantitative study of vibrational dynamics in liquids and soft materials important in different fields of chemistry.

## Methods

The experimental setup is described in detail elsewhere [24]. Here we provide a brief overview. The output of a femtosecond Ti:sapphire regenerative amplifier (Astrella, Coherent) centered at 800 nm with a repetition rate of 1 kHz, and FWHM of ~60 nm is split into three beams. The first beam (≈1 mJ/pulse) pumps a commercial TOPAS (traveling wave optical parametric amplification of superfluorescence) coupled with an NDFG (non-collinear difference frequency generation) unit to generate tunable mid-infrared (IR) pulses with FWHM of ~250 cm$^{-1}$ (spectrum of IR pulse centered at 2780 cm$^{-1}$ is shown in Fig. 1). We use a 4f pulse shaper to produce narrowband visible pulse (VIS, centered at about 800 nm, FWHM ≈ 10 cm$^{-1}$) from the second beam (≈0.8 mJ/pulse). IR and VIS beams are made collinear using a dichroic beam combiner. IR/VIS pulse pair is then routed to a displaced Sagnac interferometer to generate a local oscillator (LO) by sum frequency generation in a 10 μm thick type 1 BBO crystal. The third beam (≈1.1 mJ/pulse) is used to generate broadband terahertz (THz) pulse by two-color mixing of laser field in air plasma (spectrum of THz pulse is shown in Fig. 1). IR, VIS and LO pulses are focused by a parabolic mirror and overlapped with horizontally polarized terahertz (THz) pulse at the sample to generate third order signal. After the sample, the signal and LO are collimated with a $CaF_2$ lens (20 cm focal length) and passed through a polarizer. The signal and LO are aligned to a spectrometer (SpectraPro HRS-300, Princeton Instruments) and detected with an EMCCD detector (Newton 971, Andor). The energies of IR and VIS pulses at the sample are 1.5 μJ/pulse and 4.5 μJ/pulse, respectively. We use parallel polarization of THz, IR, VIS, and signal fields.

We measure time-domain data by varying the time delay between THz and IR/VIS pulse pair in steps of 10 fs using a motorized delay stage (V-551.2B, Physik Instrumente (PI)). The starting position of the delay stage is the same for DMSO and $SiN_x$ membrane. We scan time delay across a total time interval of 7.33 ps and 1.33 ps for DMSO and $SiN_x$, respectively. Data acquisition for one spectrum takes about 25 minutes for DMSO and 5 minutes for the $SiN_x$ membrane.

The sample cell for DMSO consists of a front and back window separated by a 1 mm thick Viton o-ring. The front window is made of a 200 μm thick silicon frame with a low-stress $SiN_x$ membrane of 50 nm thickness and 0.5 mm × 0.5 mm aperture (NX5050A, Norcada). The back window is a 2 mm thick $CaF_2$ window.

## Acknowledgements

We acknowledge the Max Planck Society for financial support.